\documentstyle[prl,aps,amsfonts,amssymb,twocolumn,epsfig]{revtex}
\begin{document}
\twocolumn[\hsize\textwidth\columnwidth\hsize\csname @twocolumnfalse\endcsname
\draft
\title{Plasma-Induced Frequency Chirp of Intense Femtosecond Lasers \\
       and Its Role in Shaping High-Order Harmonic Spectral Lines}
\author{Jung-Hoon Kim and Chang Hee Nam}
\address{Department of Physics and Coherent X-Ray Research Center,
         Korea Advanced Institute of Science and Technology,
         Taejon 305-701, Korea}
\date{\today}
\maketitle

\begin{abstract}
We investigate the self-phase modulation of intense femtosecond
laser pulses propagating in an ionizing gas and its effects on
collective properties of high-order harmonics generated in the 
medium. Plasmas produced in the medium are shown to induce a positive
frequency chirp on the leading edge of the propagating laser pulse,
which subsequently drives high harmonics to become positively 
chirped. In certain parameter regimes, the plasma-induced positive 
chirp can help to generate sharply peaked high harmonics, by 
compensating for the dynamically-induced negative chirp that is 
caused by the steep intensity profile of intense short laser pulses.
\end{abstract}


\vskip1.5pc]

Recent remarkable progress in high-power femtosecond laser 
technology has provided a novel opportunity to investigate 
high-order harmonic generation processes in an unprecedented 
high-intensity, ultrashort pulse regime. While necessary for 
generating high-order harmonics with wavelength extremely short 
compared to the laser wavelength\cite{Spiel,Chang}, the use of 
lasers in this regime has some other important merits: the harmonic
conversion efficiency is dramatically enhanced\cite{Kan1,Schn}, and
the harmonic pulse duration can be reduced to a subfemtosecond time
scale\cite{Schaf,Lee}. 

As the laser pulse evolves into the high-intensity, ultrashort 
regime, high harmonic emission begins to exhibit complicated
spectral features, such as harmonic line broadening, blueshifting, 
splitting, and smearing\cite{Schaf,Lee,Watson,Kan2,Shin,Kim1}. 
All of these features are absent in the high harmonic spectra 
observed for lasers with weak intensity well below the ionization
saturation intensity and long pulse duration (longer than a few 
hundred femtoseconds) that show well-defined narrow harmonic peaks
almost exactly at odd multiples of the fundamental laser frequency.
The involved spectral features showing up for intense short laser 
pulses can be satisfactorily accounted for, at least at the 
single-atom level, in terms of decomposition of high harmonics 
into quantum-path components, and different behaviors of the 
quantum-path components, in laser fields with rapidly-varying 
instantaneous intensity\cite{Kim1,Gaarde,Balcou}.

Since high-intensity laser pulses can ionize atoms producing 
plasmas that can affect propagating waves, a proper understanding
of macroscopic high harmonic structure necessitates a detailed 
knowledge of the plasma-related propagation effect as well as the
single-atom effect. It is known that an increase in electron 
density during the laser pulse duration causes the refractive 
index of a medium to decrease with time, which leads to frequency
upshifting and spectral broadening of the laser 
pulse\cite{Yablo,Wilks,Wood,Rae,Chessa}. Because the change 
in the driving laser spectrum should be reflected in the harmonic 
conversion, one might expect that harmonic spectral lines would 
likewise be blueshifted and broadened. In reality, however, the 
harmonic spectrum exhibits behavior more complicated than this 
expectation, as demonstrated in this paper. As a matter of fact, 
there is another source that can affect the harmonic spectral line
shape: the dynamically-induced harmonic chirp (dynamic chirp) that
is brought about by a steep pulse envelope\cite{Gaarde,Kim2}. 
The observed spectral behavior of high harmonics can only be 
understood through inspecting the temporal variation of harmonic 
frequency caused by the plasma-induced change in the laser 
frequency, in close connection with the dynamic chirp.

In this paper, we elucidate how the plasma effect modulates the 
spectrum of an intense femtosecond laser pulse, and discuss its 
subsequent influence and the effect of laser focusing on 
macroscopic high harmonic spectra. It is shown that the plasmas 
induce a positive frequency chirp on the leading edge of the laser
pulse up to the point at which a maximum blueshift is attained, 
and a negative chirp on the remaining part of the laser pulse. 
Depending on the relative amount of the plasma-induced chirp 
compared to the dynamic chirp, not only broadening but also 
narrowing can occur in high harmonic spectral lines. We 
demonstrate these using a one-dimensional (1D) model, in which 
the atomic response to the laser is calculated from the 1D 
Schr\"{o}dinger equation, and propagations of the laser and 
harmonic fields are considered in 1D space along the propagation
axis.

In order to see how the laser field $E_1$ is affected by the plasmas
(whose effects are dominated by electrons) produced in the medium, 
we begin by finding a solution of the 1D wave equation:
\begin{equation}
\frac{\partial ^2E_1(x,t)}{\partial x ^2} -\frac{1}{c^2}
\frac{\partial ^2E_1}{\partial t ^2} =
\frac{\omega ^2_p(x,t)}{c^2}E_1,
\end{equation}
where $\omega _p(x,t)= \omega _0[N_e(x,t)/N_{cr}]^{1/2}$ is the 
local plasma frequency, and $\omega _0$ is the laser frequency. The
critical plasma density $N_{cr}$ is given in Gaussian units by 
$N_{cr}=m_e\omega _0^2/4\pi e^2$, where $m_e$ is the electron mass.  
To calculate the electron density $N_e(x,t)$, we use the ADK 
model\cite{Ammo}, and consider sequential tunneling ionization up 
to as high
stages of ionization as needed, neglecting collisional ionization 
that is of little significance in the parameter regions in which
high harmonic generation experiments are commonly carried out. 
At the present gas pressures ($\leq$ 100 Torr) much lower than 1 
atm, the energy loss and temporal broadening of the laser are 
negligible\cite{Wood}; thus, we may ignore the amplitude modulation
of the laser field. Assuming the Gaussian incident pulse 
$E_1(x,t)=E_0\exp{[-(2\ln{2}/\Delta t^2) (t-x/c)^2 -iw_0(t-x/c)]}$
that is a solution to Eq.\ (1) in free space with $\Delta t$ being
the full width at half maximum (FWHM) of the pulse, we may then 
write the solution of Eq.\ (1) in the medium as
\begin{eqnarray}
E_1(x,t)&=&E_0\exp{[-(2\ln{2}/\Delta t^2)(t-x/c)^2 -i\omega_ 0t }
\nonumber \\
        & & +i\int^x n(x',t-|x-x'|/c)\frac{\omega _0}{c}dx'],
\end{eqnarray}
where $n(x,t)=[1-N_e(x,t)/N_{cr}]^{1/2}$ is the refractive index
of the medium. This expression turns out
to be a good approximate solution of Eq.\ (1) under the
conditions $N_e/N_{cr} \ll 1$ and $|c^{-1}\int^x(\partial n/
\partial t)dx'| \ll 1$, which are satisfied in the parameter 
regions considered in this paper.

To confirm that Eq.\ (2) indeed closely approximates the exact 
solution, we present in Fig.\ 1 some typical spectra of 30-fs
laser pulses at the exit of the medium (Ne gas) calculated 
from Eq.\ (2), along with those obtained from direct numerical 
calculations of Eq.\ (1). Use of the explicit expression in 
Eq.\ (2) of course significantly reduces the computational time. 
Upon comparison, it is obvious that the approximate solutions 
presented in Fig.\ 1(a) agree well with the exact numerical 
results in Fig.\ 1(b) in the parameter regions considered. It 
can be seen that, as the gas density increases and/or as the 
laser intensity increases, the spectrum shifts toward a 
higher-frequency region and becomes broader. 

More detailed features of the laser pulse passing through the 
medium can be revealed with the help of the Wigner distribution 
function\cite{Kim2}, that allows a view of temporal variation of
the laser spectrum. The Wigner distributions, calculated under 
the same conditions as in Fig.\ 1, are displayed in Fig.\ 2. It 
can be observed that, owing to the phase modulation induced by 
plasmas produced in the medium, the laser frequency increases 
with time (becomes positively chirped) in the leading edge, and 
then decreases (becomes negatively chirped) back to the original
frequency in the remaining part of the pulse. We note that at 
the moment when the plasma-induced chirp changes sign from 
positive to negative, the production rate of electrons reaches 
its maximum. This moment, at which the laser experiences a 
maximum blueshift, comes earlier in time as the laser intensity
increases, as can be seen from Fig.\ 2.  

When focused laser beams are used, as in usual high harmonic 
generation experiments, due regard should also be paid to the 
focusing effect that can change the amplitude and phase of the 
laser field. In fact, in three-dimensional (3D) simulations this 
focusing effect could have been automatically considered. 
Apparently, however, the 1D wave equation in Eq.\ (1) cannot 
deal with this effect. Nevertheless, via Eq.\ (2) the focusing 
effect can be taken into account along the propagation axis as 
follows:
\begin{equation}
E_{1f}(x,t)=E_1(x,t)f(x)\exp{[i\Phi _f(x)]},
\end{equation}
where $f(x)=[1+4(x/b)^2]^{-1/2}$ and $\Phi _f(x)=
-\tan ^{-1}(2x/b)$ represent, respectively, the amplitude and 
phase changes due to the focusing\cite{Sieg}, and $b$ is the 
confocal parameter. The laser field given in the form of Eq.\ (3)
now suitably describes the phase modulation induced by plasmas 
and the focusing effect along the axis, and can be used for
discussing the phase matching issue of high harmonics generated 
{\em on the propagation axis}.    

We next discuss the propagation of high harmonic fields. Since the 
change in the refractive index in the presence of plasmas is much
smaller for higher-frequency waves, we may ignore here the plasma 
effect. Then the Green function method enables us to write a 
solution of the 3D wave equation $\nabla ^2E_h -c^{-2}(\partial 
^2E_h/\partial t ^2) = 4\pi c^{-2}(\partial ^2P/\partial t ^2)$
in the following integral form:
\begin{equation}
E_h({\bf r},t)=-\int \frac{1}
{c^2|{\bf r}-{\bf r}'|}
\frac{\partial ^2 P({\bf r}',t')}{\partial t'^2}
d^3r',
\end{equation}
where $t'$ is the retarded time defined by 
$t'=t-|{\bf r}-{\bf r}'|/c$, and the laser-induced polarization 
$P$ is given in terms of the gas density $N_0$ and the atomic 
dipole moment $d$ by $P({\bf r}',t')=N_0d({\bf r}',t')$.
Equation (4) clearly indicates that the harmonic field $E_h$ at 
$t$ is determined by the coherent sum of the dipole accelerations
of atoms in the medium calculated at the retarded time $t'$.

In our 1D calculations, a medium lies on the propagation axis, 
and the integration in Eq.\ (4) is performed in practice over
the 1D space along the axis. The medium, a Ne gas of length 
$l= 700$ $\mu$m at $\simeq$ 28 Torr, is uniformly discretized by
200 points, and at each point the dipole acceleration is 
calculated by numerically solving the 1D Schr\"{o}dinger equation
for an atom in the laser field $E_{1f}$ given in Eq.\ (3), with 
$\Delta t=30$ fs, $\lambda = 800$ nm, $b=4$ mm, and $I=1\times 
10^{15}$ W/cm$^2$ at the entrance of the medium. After suitably 
weighted by a constant factor to yield the correct gas density, 
the results are then added according to Eq.\ (4) to give $E_h$. 

The high harmonic spectrum calculated in the above way is 
presented in Fig.\ 3, which shows the respective roles played
by the laser focusing and plasmas in the formation of macroscopic
harmonic fields. In Fig.\ 3(b), we neglect the plasma effect by 
setting $n(x,t)=1$ to concentrate only on the focusing effect. 
Both the focusing and plasma effects are fully considered in 
Fig.\ 3(c), and for comparison we present in Fig.\ 3(a) a 
single-atom spectrum calculated for an atom located at the 
entrance of the medium. Whereas the single-atom spectrum in 
Fig.\ 3(a) is smeared with a complicated structure in the plateau
region, the macroscopic harmonic spectrum in Fig.\ 3(b) exhibits 
discrete harmonic peaks. This is because the intensity-sensitive 
harmonic components (the long quantum-path component and 
multiple-recollision components), which give rise to the 
complicated structure, are suppressed due to their poor 
phase-matching conditions, leaving only the short quantum-path 
component\cite{Kim1,Sali}. Here we emphasize that the variation 
in the laser intensity caused by the focusing as represented by 
$f^2(x)$ is mainly responsible for this cleaning up of the 
harmonic spectrum. At very low gas densities, we may observe a 
harmonic spectrum similar to that shown in Fig.\ 3(b). However, 
at the gas density used in Fig.\ 3(c), a further change in the 
harmonic spectrum is caused by the plasma effect: harmonics of 
low order (43rd and below) undergo  spectral broadening, while 
higher-order harmonics get sharpened. 

The plasma-induced harmonic line broadening and narrowing are 
detailed in Figs.\ 4 and 5, respectively.  The mechanism that 
shapes high harmonic lines can be explained by making a 
comparison between the plasma-induced chirp and the 
dynamically-induced chirp\cite{Schaf,Gaarde,Kim2}. Considering 
that only the leading edge of the laser pulse, where the 
depletion of neutral atoms is not severe and the electron density
is low, is important for the phase-matched harmonic generation, 
we focus only on this part of the laser pulse. The leading edge 
of the laser pulse, in which the laser frequency increases with 
time due to the plasma effect (Fig.\ 2), tends to generate 
positively chirped high harmonics. This effect is opposed by the 
dynamic single-atom effect that, in the present case, forces high
harmonics to become negatively chirped\cite{note}. Depending on 
system parameters and harmonic order, one effect matches or 
dominates over the other. For instance, for the low-order 
harmonics in Fig.\ 4, the dynamic negative chirp is very small 
[Fig.\ 4(b)], and the plasma effect dominantly affects the 
time-frequency characteristics of high harmonics in such a way 
as to make high harmonics become positively chirped [Fig.\ 4(d)], 
leading to spectral line broadening [Fig.\ 4(c)]. In contrast, 
the dynamic negative chirp of higher-order harmonics in Fig.\ 5(b)
is substantially large, and the plasma-induced positive chirp 
more or less cancels out the dynamic chirp [Fig.\ 5(d)], resulting
in harmonic line narrowing [Fig.\ 5(c)]. In both cases of Figs.\ 4
and 5, the plasma effect blueshifts the central frequencies of 
high harmonics.

In conclusion, we have investigated the plasma-induced phase 
modulation of intense femtosecond laser pulses and its effects on
macroscopic high harmonic spectra, using a simple but accurate 1D
model. It has been shown that the plasmas produced in the medium 
induce a positive frequency chirp on the leading edge of the laser 
pulses. Strikingly, the plasma-induced chirp, which broadens the 
laser spectrum, can lead not only to broadening but also to 
narrowing of harmonic spectral lines. The underlying mechanism
has been explained by comparing the plasma-induced positive chirp 
with the dynamically-induced negative chirp. In stark contrast to 
the widely reported finding (in regimes well below saturation)
that high harmonics are negatively 
chirped\cite{Sali,Altu,Seki}, this study clearly 
demonstrates that, under laser fields of high intensity and gas 
pressures not too low, high harmonics can 
become free of chirp, or even become
positively chirped, thanks to the plasma effect. 
This suggests a way to control the chirp and spectral line shape 
of high harmonics using the plasma effect. 

This work was supported by the Ministry of Science and Technology 
of Korea through the Creative Research Initiative Program.

\begin{figure}[h]
\centering
\epsfig{width=0.95 \linewidth,file=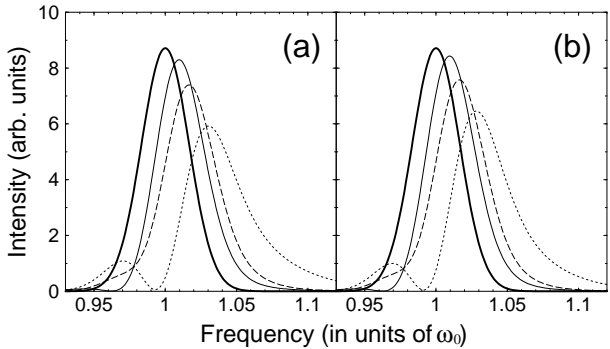}
\caption[ ]{
(a) Laser spectra obtained from Eq.\ (2) for a 30-fs (FWHM) laser pulse
of wavelength $\lambda =800$ nm after propagating through a Ne gas medium of
length $l=700$ $\mu$m.
The laser intensity and gas density are given by
$I=1\times 10^{15}$ W/cm$^2$ and $N_0=1\times 10^{18}$ cm$^{-3}$ ($\simeq$
28 Torr) (thin solid line);
$I=1\times 10^{15}$ W/cm$^2$ and $N_0=3\times 10^{18}$ cm$^{-3}$
($\simeq$ 85 Torr) (dotted line);
$I=3\times 10^{15}$ W/cm$^2$ and $N_0=1\times 10^{18}$ cm$^{-3}$
(dashed line). The spectrum of the incident laser pulse is drawn by
a thick solid line.
(b) Laser spectra obtained by direct numerical calculations of Eq.\ (1)
for the same parameters as in (a).}
\label{fig1}
\end{figure}

\begin{figure}[h]
\centering
\epsfig{width=0.95 \linewidth,file=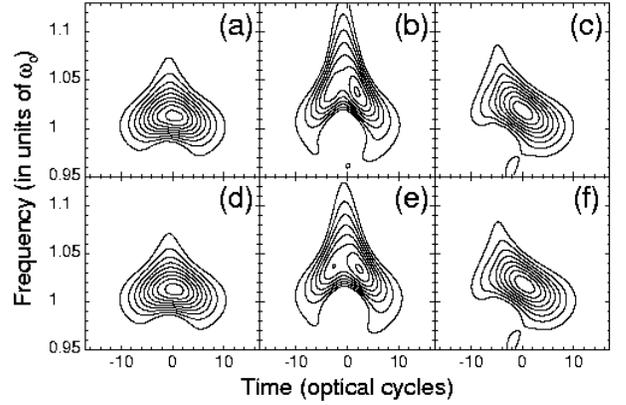}
\caption[ ]{
Wigner distributions of the laser pulses specified in Fig.\ 1.
In (a)-(c) [(d)-(f)], the same parameters are used as for the thin solid,
dotted and dashed lines in Fig.\ 1(a) [Fig.\ 1(b)], respectively.
Only positive contour lines are shown.
}\label{fig2}
\end{figure}

\begin{figure}[h]
\centering
\epsfig{width=0.95 \linewidth,file=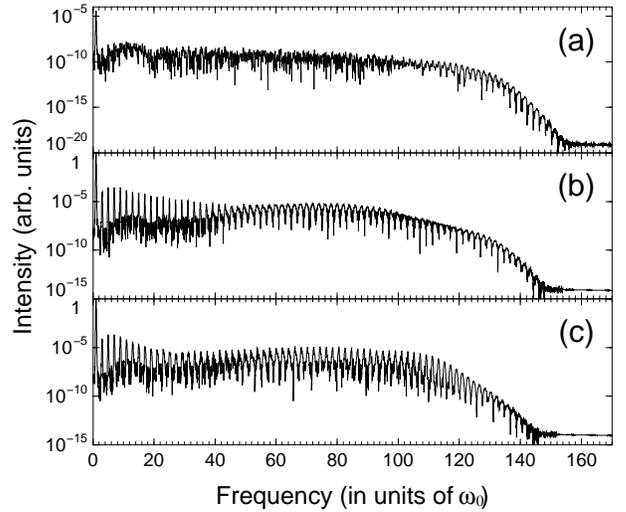}
\caption[ ]{
(a) Single-atom harmonic spectrum for a Ne atom placed at the
entrance of the medium.
(b) Macroscopic harmonic spectrum (in the far field, on the axis)
considering only the effect of laser focusing.
(c) Macroscopic harmonic spectrum taking into account both
the focusing and plasma effects.
A Ne gas medium of density $N_0=1\times 10^{18}$ cm$^{-3}$
and length $l=700$ $\mu$m is irradiated by a 30-fs (FWHM) Gaussian laser pulse
of wavelength $\lambda =800$ nm.
The medium is centered at $x=2$ mm behind the focus,
and because of focusing ($b=4$ mm) the laser intensity decreases
from $I=1\times 10^{15}$ W/cm$^2$ at the entrance
to $I=0.7\times 10^{15}$ W/cm$^2$ at the exit of the medium.
}\label{fig3}
\end{figure}

\begin{figure}[h]
\centering
\epsfig{width=0.95 \linewidth,file=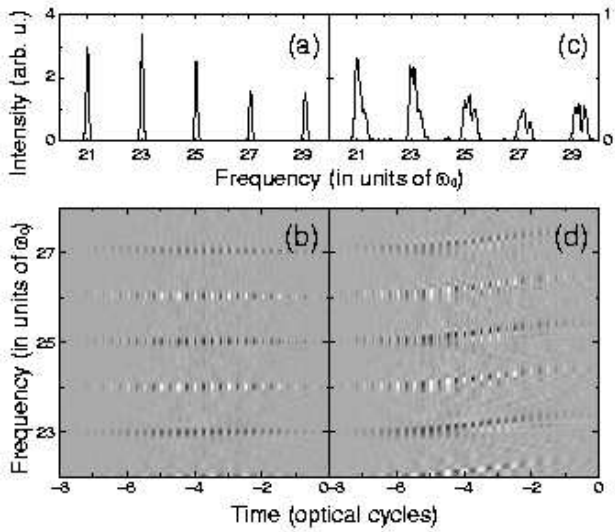}
\caption[ ]{
(a) Enlarged view (on a linear scale) of Fig.\ 3(b) between $20\omega _0$
and $30\omega _0$ (only the focusing effect is considered).
(b) Wigner distribution of the harmonics chosen in (a).
(c) [(d)] Same as (a) [(b)] except that harmonics are
chosen from Fig.\ 3(c) (both the focusing and plasma effects are
considered). Positive and negative values of the Wigner distribution
are colored black and white, respectively.
}\label{fig4}
\end{figure}

\begin{figure}[h]
\centering
\epsfig{width=0.95 \linewidth,file=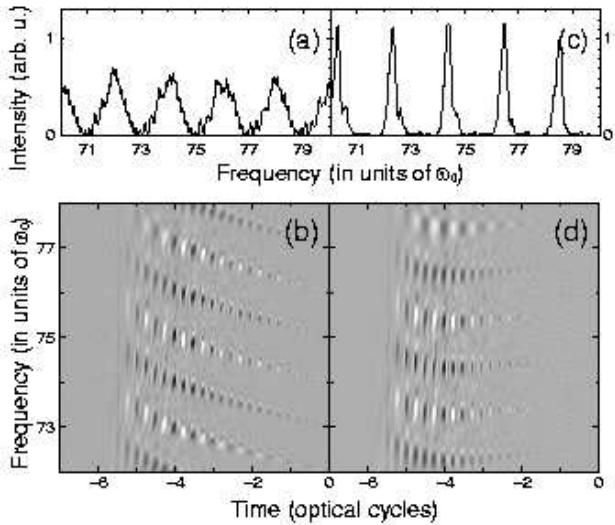}
\caption[ ]{
Same as Fig.\ 4 except that harmonics between $70\omega _0$ and $80\omega _0$
are chosen here.
}\label{fig5}
\end{figure}


\begin{references}
\bibitem{note} The laser pulse experienced by atoms in the case of 
         Fig.\ 3(b) resembles in form that used in our previous 
         discussions on the coherent sum method\cite{Kim1} and the
         time-frequency analysis\cite{Kim2}. Thus, the slope of 
         the Wigner distributions shown in Figs.\ 4(b) and 5(b)
         gives the dynamic chirp.
\end{references}
\end{document}